\documentclass[twocolumn,showpacs,preprintnumbers,amsmath,amssymb,epsfig]{revtex4}

\usepackage{epsfig}

\newcommand{\be}{\begin{eqnarray}}
\newcommand{\ee}{\end{eqnarray}}

\begin{document}

\title{Factorization of Numbers with the temporal Talbot effect: Optical implementation by a sequence of shaped ultrashort pulses}
\author{Damien Bigourd$^1$,  B\'eatrice Chatel$^{1**}$, Wolfgang P. Schleich$^2$ and Bertrand Girard$^{1*}$}
\affiliation{$^1$ Laboratoire de Collisions, Agr\'egats, R\'eactivit\'e (UMR 5589, CNRS - Universit\'e de Toulouse, UPS), IRSAMC, Toulouse, France
 \\ $^2$Institut f\"{u}r Quantenphysik,
Universit\"{a}t Ulm, Albert-Einstein-Allee 11, D-89081 Ulm, Germany}

\date{\today}

\begin{abstract}

We report on the successful operation of an analogue computer designed to factor numbers.
Our device relies solely on the interference of classical light and brings together the field
of ultrashort laser pulses with number theory. Indeed, the frequency component of the electric
 field corresponding to a sequence of appropriately shaped femtosecond pulses is determined by
  a Gauss sum which allows us to find the factors of a number.

\end{abstract}

\maketitle

In 1836 Henry Fox Talbot used "a magnifying glass of considerable power" \cite{Talbot1836}
to investigate the interference pattern of light emerging from a diffraction grating produced
 by Joseph Fraunhofer. Talbot noticed "a curious effect": The interference patterns in planes
 parallel to the grating repeated themselves periodically as the distance between the plane and
  the grating increased. Almost fifty years later this self-imaging effect was rediscovered and
  explained by Lord Rayleigh \cite{Rayleigh1881}. Today the Talbot effect \cite{SchleichBerry01}
   manifests itself not only in electromagnetic waves \cite{Patorski1989} but also in matter waves
    with applications ranging from the observation of interferences in $\textrm{C}_{60}$ molecules
    \cite{Zeilinger02_mol_int} to lithography \cite{Pfau97}. In the present paper we report on a modern
     day variant of the Talbot effect using appropriately shaped femtosecond laser pulses and use it to factor numbers.

Factorizing numbers is an important problem in network as well as
security systems \cite{Koblitz94}. Many attempts have been made to use quantum
systems to dramatically increase the efficiency. However, even today
the challenge remains and only small numbers  \cite{Chuang01} have been factorized
using a quantum algorithm.

At the same time, other physics-based methods for factorizing
numbers have been proposed
\cite{Merkel06FP}
. One of them relies
on the properties of the truncated Gauss sum
 \be\label{gaussum}
 {\cal A}_N^{(M)} \left( l \right) = \frac{1}{{M + 1}}\sum\limits_{m = 0}^M
{\exp \left( { -2\pi i m^2 \frac{N}{l}} \right)} \ee
 consisting of
 $M+1$ terms and $N$ is the number to be factored. The argument $l$
 scans through all integers between $2$
 and $\sqrt{N}$ for possible factors. When $l$ is not a factor, the quadratic phases oscillate rapidly with $m$ and the sum takes on small values.
When $l$ is a factor, then all the phases are multiple of $2\pi$ and
the sum is equal to unity.

The proposed implementations of $ {\cal A}_N^{(M)}$  are based on multipath interferences \cite{Merkel06FP}. Each path produces one term in the Gauss sum. The difficulty is to find a system which is experimentally accessible and in which the required phase in Eq. (\ref{gaussum}) is obtained by a simple variation of a physical parameter.

So far this strict condition has not yet been fulfilled. Nevertheless, several experiments in which each phase of the Gauss sum is separately computed have recently succeeded to demonstrate the ability of Gauss sums to factorize numbers with physical systems. In two experiments based on NMR techniques \cite{MehringPRL07,
 MaheshPRA07} the nuclear spins are driven by a series of radio-frequency pulses. In a more recent experiment, cold atoms are excited by a sequence of Raman $\pi$-pulses \cite{ErtmerSchleich_factorisation-Rb07}.

In the present paper we introduce an all optical approach towards
factoring numbers relying on modern pulse shaping technology.
Indeed, the generation of arbitrarily shaped optical waveforms
\cite{weiner00} is of great interest in a number of fields ranging
from coherent control \cite{silberberg98}
to information processing \cite{BucksbaumScience00, Weiner00SHG,
AMitay02ANDgate}. For example, pulse shapers have led to an elegant
implementation of the Grover search algorithm using Rydberg atoms as
quantum registers \cite{BucksbaumScience00}. Moreover, optical
realizations of the Grover \cite{Spreeuw_PRL02} or the
Bernstein-Vazirani  \cite{Massar_PRL03, Walmsley_PRA04} algorithms
have been used. Our work extends this line of research to factoring
numbers using the Talbot effect.

Three elements determine the Talbot effect: (i) a grating which is periodic in space and creates a periodic spatial field distribution, (ii) interference of the waves emerging from each slit of the grating and (iii) the paraxial approximation of classical optics which leads to the accumulation of quadratic phases in the time evolution of these waves. As a consequence, the intensity distribution of light on a screen is determined by a Gauss sum.

The present implementation follows exactly this recipe except that it takes place in the time rather than the space domain. For this temporal Talbot effect \cite{Morgner}, we
consider an electric field
\be\label{fieldt}
\tilde E \left(t \right)= \sum\limits_{m }^{} w_m \, {e^{i \theta _m } \, e^{-i \omega _L t} \, \delta \left( t-\tau_m \right)}
\ee
consisting of a sequence of short pulses approximated by delta functions. The pulses of carrier frequency $\omega_L$ and phases $\theta_m$ appear at times $\tau_m$ and the weight factors $w_m $ guarantee that the energy of the pulse remains finite.

The frequency component
\be \label{TF}
E\left( \omega \right)= \int_{ - \infty }^\infty  {dt\, \tilde E(t)\,e^{i\omega t} },
\ee
of this pulse sequence defined by the Fourier transform follows from the interference of the Fourier components of  the individual pulses
\be\label{fieldw} E\left( \Delta \omega \right) = \sum\limits_{m} w_m  {\exp {\left[ i\left(
{\theta _m  + \tau_m  \, \Delta \omega } \right) \right] } } \ee
with $\Delta \omega =\omega-\omega_L$.

By imprinting appropriate phases on the pulses with pulse shapers we can obtain the quadratic phases characteristic of the Talbot effect. For example, the special choice
$\theta_m \equiv - 2\pi m^{2} N/l $ with times $\tau_m \equiv m T$ and the weight function
\be
w_m =
\left\{ {\begin{array}{*{20}c}
   {{(M + 1)}^{-1} \,\, {\rm{ for }} \,\, 0 \le m \le M} \hfill  \\
   {0 \,\, {\rm{ otherwise }}} \hfill  \\
\end{array}} \right.,
\ee yields for $\Delta \omega =0$, that is $\omega=\omega_L$, the Gauss sum Eq. (\ref{gaussum}). This Gauss sum is thus directly calculated by multipath optical interferences between the optical pulses.

The laser system is a conventional Ti: Sapphire laser
 delivering pulses of $\tau_L = 30\, {\rm fs}$ at $805 \, {\rm nm}$ with $80$
MHz repetition rate. The laser pulses are shaped with a
programmable 640 pixels phase and amplitude pulse-shaper offering a shaping window of $T_w \simeq 30 \, {\rm ps}$ \cite{pulseshaperRSI04}.

In order to generate at once the shaped pulse sequence required by Eq. (\ref{fieldt}), the complex spectral mask
\be\label{mask} H_\theta \left( \omega  \right) = w_m \sum\limits_{m = 0}^M {\exp {\left[ {i\left( {\theta _m  +  \tau_m \, \Delta \omega  } \right)}\right] }} \ee
is applied with the pulse shaper to modify the Fourier Transform limited input laser pulse: $E_{out} \left(\omega  \right)  = H_\theta \left(\omega  \right) E_{in} \left(\omega  \right) $.
Each term of the sum in Eq. (\ref{mask}) is therefore produced by one ultrashort pulse delayed by $\tau_m$ and with an extra phase shift $\theta_m$. Here we
choose $T=200 \, {\rm fs}$ in order to produce a sequence of well separated pulses.

 The interference produced by the pulse sequence is simply analyzed with a high resolution spectrometer. We measure the spectral intensity at the central wavelength $\lambda_L = 2\pi\, c / \omega_L $ and thus retrieve the Gauss sum for each
$l$. The experiment is performed for $l$ ranging between $2$ and $\sqrt{N}$ in order to discriminate
factors from non-factors.
\begin{figure}[!ht]
\begin{center}
\epsfig{figure=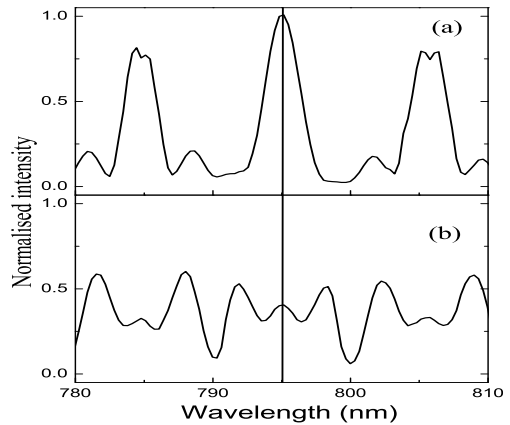,width=0.85\linewidth, height=6cm}
\caption[spectrum]{Spectra of a $M+1=4$ pulses sequence ($N=105$).  $l= 3$
(a) and $l= 9$ (b). The vertical line represents $\omega_L$.}
\label{spectrum}
\end{center}
\end{figure}

Figure \ref{spectrum} shows two typical spectra obtained for $N=
105$ and $l= 3 \; {\rm or } \;9$ with $M+1=4$ pulses and after averaging over ca 10 s. The complex structure reflects the multipulse
interferences and underlines  the requested high resolution (about $0.06$ nm). The normalization is obtained from full and minimal shaper transmission.

\begin{figure}[!ht]
\begin{center}
\epsfig{figure=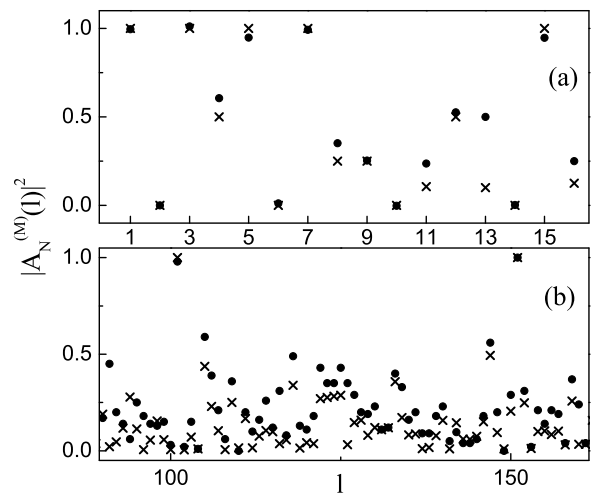,width=\linewidth}
\caption[sum]{Experimental realization of factoring using a sequence
of shaped ultrashort pulses: (a) $N= 105 =3*5*7$ with 4 pulses,
 (b)
 $N=15251=101*151$ with $M+1=9$ pulses. Experiment (dots); Theory (crosses).} \label{sum}
\end{center}
\end{figure}

Several
numbers have been factorized with this method. In Figure \ref{sum} we display the
results of our optical implementation of the factorization scheme
based on Gauss sum for $N= 105 =3*5*7$ obtained with a four pulse sequence
(a),
and for $N=15251=101*151$ with a nine pulse sequence (b). The first
example consists of the product of twin primes whereas the second
consist of quite far primes allowing to test the validity of the
method.

The experimental data indicated by black dots are compared with the
expected values $\left| {\cal A}_N^{(M)} \left( l \right) \right|^2
$ depicted by crosses and the agreement is very good, particularly
for the factors whose Gauss sum comes out very clearly. The
experimental contrast is in general smaller than expected. This
reduction could be due to several experimental limitations :  (i)
Our shaper is pixellated in the spectral domain and therefore
introduces temporal replica. These replica are separated by $ 35\,
{\rm ps}$ and are particularly broad and weak due to the nonlinear
dispersion in the mask plane \cite{wang_shaping2001,
Vaughan_optexp2006} which was carefully calibrated. This time window
of $ 35\, {\rm ps}$ is restricted down to $ 28\, {\rm ps}$ by the
effect of the gaussian envelope due to the spatial beam profile in
the mask plane \cite{pulseshaperRSI04, Vaughan_optexp2006}. Its
consequences are limited here by working on only a fraction (3 to 7
ps) of the shaping window.
(ii) Another consequence of pixellation is the hole in amplitude
associated to large phase steps between consecutive pixels
\cite{RbShapingAPB04} which may induce small distortions as compared
to the ideal transmission $H_\theta (\omega)$.
(iii) The main limitation to the extinction ratio (currently of
20\,dB) is due to the gaps between pixels in the LCD (3\% of the
pixel width) adding a non programmable pulse at $t=0$, which
participates also to this loss of contrast. This contribution is
difficult to compensate and produces undesired interferences with
the pulse train\cite{pulseshaperRSI04}. (iv) Finally the resolutions
of both pulse shaper and spectrometer limit the ultimate contrast
which can be achieved. Both are carefully calibrated following the procedure described in \cite{pulseshaperRSI04}.

 A key issue in the efficiency and
reliability of this scheme is the choice of the truncation parameter
$M$ of the Gauss sum. This question is closely related to the
phenomenon of ghosts factors \cite{Schleich07truncation}. Indeed,
for certain integer arguments $l$, the Gauss sum can take values
close to unity even when $l$ is not a factor of N. Ghosts can be
suppressed  \cite{Schleich07truncation} below the threshold of $1/
\sqrt{2}$ by choosing
$M\simeq 0.7\sqrt[4]{N}$.

The example $N= 19043=p(p+2)$ with $p=137$ is perfectly suited to
test the predictions of Ref. \cite{Schleich07truncation} concerning
ghost factors. In this case, $N$ consists of the product of twin
primes which are approximately equal and of the order of $\sqrt{N}
\cong 137.996$. In this way we can test our method at the upper
boundary $\sqrt{N}$ of our set of trial factors. For this purpose we
first note that for any integer number $N$ consisting of the product
of twin primes the elementary relation $N=p(p+2)=(p+1)^2 - 1$ yields
the approximation $p+1 \cong \sqrt{N}$ together with the
decomposition $ {N} / {(p+1)} = (p+1) - {1}/{(p+1)}$. As a result
the truncated Gauss sum reduces to \be\label{gaussumtronc}
 {\cal A}_{p(p+2)}^{(M)} \left( p+1 \right) = \frac{1}{{M + 1}}\sum\limits_{m = 0}^M
{\exp \left( { -2\pi i  \frac{m^2}{p+1}} \right)} \ee Since $1 \ll
M$ and $1 \ll (p+1)$ we can approximate this sum by a Fresnel
integral which yields \cite{Schleich07truncation} the scaling
$M\propto\sqrt{p+1} \cong \sqrt[4]{N}$.

\begin{figure}[!ht]
\begin{center}
\epsfig{figure=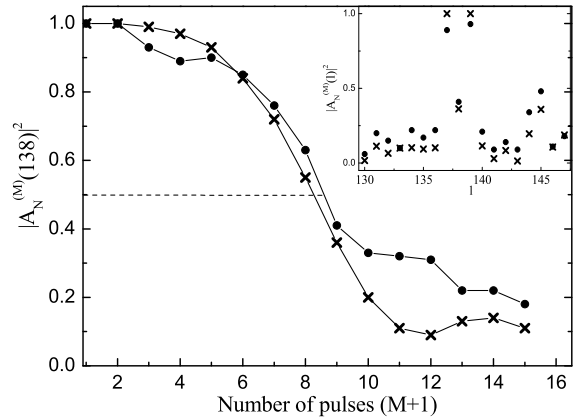,width=1\linewidth}
\caption[contrast]
{Supression of the ghost factor $l=138$ of $N=19043=137*139$ for increasing number of pulses. 
Insert : Experimental realization of factoring  $N= 19043$ with $M+1=9$
pulses. Experiment (dots); Theory (crosses).} \label{contrast}
\end{center}
\end{figure}

In Fig. \ref{contrast} we display by crosses the exact sum $\left|
{\cal A}_{19043}^{(M)} \left( 138 \right) \right|^2$ as a function
of $M$. We note the slow decay and the oscillations due to the
Fresnel integral. Solid dots representing our measurements follow
this behavior. The general trend is well reproduced. However the
experimental uncertainties do not allow to reproduce fully the
expected oscillations. Moreover, we find the predicted threshold
$M\simeq 0.7\sqrt[4]{19043}\simeq 8$. In the insert, an experimental
realization of factoring $N= 19043 =137*139$ with a 9 pulses
sequence is shown as an example. Theory and experiments are also in
excellent agreement.


Our work clearly demonstrates that we can use shaped femtosecond pulses to implement Gauss sums and factor numbers.
However, many generalizations offer themselves: (i) So far we have only made use of the phases $\theta_m$ in the frequency representation Eq. (\ref{fieldw}) of the electric field. The second contribution to the phase, that is the product $\tau_m  \, \Delta \omega $ did not enter since we set $\Delta \omega =0$. (ii) Since we have only the single parameter $\theta_m$ at our disposal, the number $N$ to be factored and the trial factor $l$ cannot be varied independently. (iii) Finally we have pursued a sequential rather than a parallel approach. Indeed, we have only used a single spectral component.

The activation of the so far unused phase $\tau_m  \, \Delta \omega $  solves all three problems. Since now we have two parameters we can encode $N$ in $\tau_m$ and $l$ in $\Delta \omega $. By recording the complete spectrum we achieve a massive parallelism.


The choice of $\theta_m = 0$  and $\tau_m = 2\pi m^2N \alpha$ with
the numerical constant $\alpha$ also yields the Gauss sum $ {\cal
A}_N^{(M)} $ and illustrates this new approach. Here the spacing
between pulses increases quadratically and $l$ is inversely
proportional to $\Delta \omega $ such that a single spectrum
directly contains all the information.

However, some remaining difficulties need to be overcome: (i) The Gaussian shape of the spectral profile leads to ponderations in the Gauss terms which have to
be taken into account. (ii) The variation of $l$ between 2 and $\sqrt{N}$ {\it i. e.} on several orders of magnitudes puts severe constraints on the spectral resolution necessary to carry the experiment.
(iii) Finally, the number of pulses is limited by $\sqrt{T_w/3\tau_L} \simeq 10$ with our present set-up.

The quadratic spacing of the pulses required in the above approach might represent a severe problem.
The choice $\theta_m = 0 $ and $\tau_m \equiv 2 \pi (m + m^2 /N )$ which leads to the Gauss sum \cite{Merkel06FP}
 \be\label{gaussum_reciproc}
 {\cal S}_N(\Delta \omega)  = \sum\limits_{m}
{w_m  \exp \left[ { 2\pi i(m + \frac {m^2}{N} ) \Delta \omega} \right]} \ee
might be an interesting way around, since it also allows us to factor numbers. In contrast to the truncated Gauss sum ${\cal A}_N^{(M)} $ which only needs to be recorded at integer arguments, the sum ${\cal S}_N$ relies on a continuous argument. Here we need the complete spectrum. In return ${\cal S}_N$ displays interesting scaling properties which enable us to use the interference pattern for $N$ to factor the number $N'$ by rescaling the frequency axis \cite{Merkel07PRA2}.

We conclude by noting that our implementation is closely connected to the Talbot effect in a harmonic oscillator \cite{Agarwal95}. Indeed, an initial wave function consisting of an array of sharp maxima located at integer multiples $m$ of a period $d$ accumulates quadratic phases in its time evolution. On first sight this behavior is surprising since the energy spectrum of the harmonic oscillator is linear. However, due to the quadratic dependence of the energy on the position, the maxima at $md$ translate into quadratic phases  $(md)^2$. A similar behavior was noted \cite{Schleich00} in the quantum carpets woven by a wave packet moving in a harmonic oscillator and consisting only on energy eigenstates with quadratic quantum numbers.

Although the electromagnetic field represents a harmonic oscillator it might be easier to realize a factorization scheme based on this effect using a mechanical oscillator. A laser cooled atom into an optical lattice and prepared in its motional ground state offers a possible realization. An absorption grating \cite{Schleich05} produced by a standing light wave can prepare the periodic array of narrow wave packets. Moreover, the coupling of the center-of-mass motion to a quantized standing light field \cite{Factor2} can introduce entanglement into the Talbot effect making factorization with entangled Gauss sums viable.

In summary, we have factorized numbers through the implementation of a Gauss sum with optical interferences produced by a sequence of shaped short laser pulses. This work opens the route to further promising developments based on the wide flexibility offered by optical interferences.

E. Baynard and S. Faure are acknowledged for their technical help. We enjoyed fruitful discussions with G.S. Agarwal, A. Monmayrant and I. Walmsley.
This work has been supported by the Agence Nationale de la Recherche (Contract ANR - 06-BLAN-0004) and the Del Duca foundation. WPS is grateful to the Max-Planck Society and the A. von Humboldt Stiftung for their support. Moreover, he acknowledges the support by the Ministerium f\"{u}r Wissenschaft und Kunst, Baden-W\"{u}rttemberg and the Landesstiftung Baden-W\"urttemberg in the framework of the Quantum Information Highway A8 and the Center for Quantum Engineering.

$^*$ Member of the Institut Universitaire de France
$^{**}$\textit{corresponding author: beatrice@irsamc.ups-tlse.fr}

\begin{thebibliography}{33}
\expandafter\ifx\csname
natexlab\endcsname\relax\def\natexlab#1{#1}\fi
\expandafter\ifx\csname bibnamefont\endcsname\relax
  \def\bibnamefont#1{#1}\fi
\expandafter\ifx\csname bibfnamefont\endcsname\relax
  \def\bibfnamefont#1{#1}\fi
\expandafter\ifx\csname citenamefont\endcsname\relax
  \def\citenamefont#1{#1}\fi
\expandafter\ifx\csname url\endcsname\relax
  \def\url#1{\texttt{#1}}\fi
\expandafter\ifx\csname urlprefix\endcsname\relax\def\urlprefix{URL
}\fi \providecommand{\bibinfo}[2]{#2}
\providecommand{\eprint}[2][]{\url{#2}}

\bibitem[{\citenamefont{Talbot}(1836)}]{Talbot1836}
\bibinfo{author}{\bibfnamefont{H.~F.} \bibnamefont{Talbot}},
  \bibinfo{journal}{Philos. Mag.} \textbf{\bibinfo{volume}{9}},
  \bibinfo{pages}{401} (\bibinfo{year}{1836}).

\bibitem[{\citenamefont{Rayleigh}(1881)}]{Rayleigh1881}
\bibinfo{author}{\bibfnamefont{L.}~\bibnamefont{Rayleigh}},
  \bibinfo{journal}{Philos. Mag.} \textbf{\bibinfo{volume}{11}},
  \bibinfo{pages}{196} (\bibinfo{year}{1881}).

\bibitem[{\citenamefont{Berry {\it et~al.}}(2001)\citenamefont{Berry, Marzoli,
  and Schleich}}]{SchleichBerry01}
\bibinfo{author}{\bibfnamefont{M.~J.} \bibnamefont{Berry}} \bibnamefont{{\it
  et~al.}}, \bibinfo{journal}{Physics World} \textbf{\bibinfo{volume}{14}},
  \bibinfo{pages}{39} (\bibinfo{year}{2001}).

\bibitem[{\citenamefont{Patorski}(1989)}]{Patorski1989}
\bibinfo{author}{\bibfnamefont{K.}~\bibnamefont{Patorski}}, in
  \emph{\bibinfo{booktitle}{Progress in Optics}}, edited by
  \bibinfo{editor}{\bibfnamefont{E.}~\bibnamefont{Wolf}}
  (\bibinfo{publisher}{North-Holland}, \bibinfo{address}{Amsterdam},
  \bibinfo{year}{1989}), vol.~\bibinfo{volume}{28}, p.~\bibinfo{pages}{1}.

\bibitem[{\citenamefont{Brezger {\it et~al.}}(2002)\citenamefont{Brezger,
  Hackermuller, Uttenthaler, Petschinka, Arndt, and
  Zeilinger}}]{Zeilinger02_mol_int}
\bibinfo{author}{\bibfnamefont{B.}~\bibnamefont{Brezger}} \bibnamefont{{\it
  et~al.}}, \bibinfo{journal}{Phys. Rev. Lett.} \textbf{\bibinfo{volume}{88}},
  \bibinfo{pages}{100404} (\bibinfo{year}{2002}).

\bibitem[{\citenamefont{Nowak {\it et~al.}}(1997)\citenamefont{Nowak,
  Kurtsiefer, Pfau, and David}}]{Pfau97}
\bibinfo{author}{\bibfnamefont{S.}~\bibnamefont{Nowak}} \bibnamefont{{\it
  et~al.}}, \bibinfo{journal}{Opt. Lett.} \textbf{\bibinfo{volume}{22}},
  \bibinfo{pages}{1430} (\bibinfo{year}{1997}).

\bibitem[{\citenamefont{Koblitz}(1994)}]{Koblitz94}
\bibinfo{author}{\bibfnamefont{N.}~\bibnamefont{Koblitz}},
  \emph{\bibinfo{title}{A Course in Number Theory and Cryptography}}
  (\bibinfo{publisher}{Springer}, \bibinfo{address}{New York},
  \bibinfo{year}{1994}).

\bibitem[{\citenamefont{Vandersypen {\it
  et~al.}}(2001)\citenamefont{Vandersypen, Steffen, Breyta, Yannoni, Sherwood,
  and Chuang}}]{Chuang01}
\bibinfo{author}{\bibfnamefont{L.~M.~K.} \bibnamefont{Vandersypen}}
  \bibnamefont{{\it et~al.}}, \bibinfo{journal}{Nature}
  \textbf{\bibinfo{volume}{414}}, \bibinfo{pages}{883} (\bibinfo{year}{2001}).

\bibitem[{\citenamefont{Merkel {\it
  et~al.}}(2006{\natexlab{a}})\citenamefont{Merkel, Sh.~Averbukh, Girard,
  Paulus, and Schleich}}]{Merkel06FP}
\bibinfo{author}{\bibfnamefont{W.}~\bibnamefont{Merkel}} \bibnamefont{{\it
  et~al.}}, \bibinfo{journal}{Fortschr. Phys.} \textbf{\bibinfo{volume}{54}},
  \bibinfo{pages}{856} (\bibinfo{year}{2006}{\natexlab{a}});
 \bibinfo{journal}{Int. J. Mod. Phys. B}
  \textbf{\bibinfo{volume}{20}}, \bibinfo{pages}{1893}
  (\bibinfo{year}{2006}{\natexlab{b}});
\bibinfo{journal}{Phys. Rev. A} \textbf{\bibinfo{volume}{76}}, \bibinfo{pages}{023417}, (\bibinfo{year}{2007}{\natexlab{a}}).
\bibinfo{author}{\bibfnamefont{J.}~\bibnamefont{Clauser}} \bibnamefont{{\it
  et~al.}}, \bibinfo{journal}{Phys. Rev. A} \textbf{\bibinfo{volume}{53}},
  \bibinfo{pages}{4587} (\bibinfo{year}{1996}).
\bibinfo{author}{\bibfnamefont{W.}~\bibnamefont{Harter}},
  \bibinfo{journal}{Phys. Rev. A} \textbf{\bibinfo{volume}{64}},
  \bibinfo{pages}{012312} (\bibinfo{year}{2001}).

\bibitem[{\citenamefont{Mehring {\it et~al.}}(2007)\citenamefont{Mehring,
  Mueller, Sh.~Averbukh, Merkel, and Schleich}}]{MehringPRL07}
\bibinfo{author}{\bibfnamefont{M.}~\bibnamefont{Mehring}} \bibnamefont{{\it
  et~al.}}, \bibinfo{journal}{Phys. Rev. Lett.} \textbf{\bibinfo{volume}{98}},
  \bibinfo{pages}{120502} (\bibinfo{year}{2007}).

\bibitem[{\citenamefont{Mahesh {\it et~al.}}(2007)\citenamefont{Mahesh,
  Rajendran, Peng, and Suter}}]{MaheshPRA07}
\bibinfo{author}{\bibfnamefont{T.~S.} \bibnamefont{Mahesh}} \bibnamefont{{\it
  et~al.}}, \bibinfo{journal}{Phys. Rev. A} \textbf{\bibinfo{volume}{75}},
  \bibinfo{pages}{062303} (\bibinfo{year}{2007}).

\bibitem[{\citenamefont{Gilowski {\it et~al.}}(2007)\citenamefont{Gilowski,
  Wendrich, Muller, Jentsch, Ertmer, Rasel, and
  Schleich}}]{ErtmerSchleich_factorisation-Rb07}
\bibinfo{author}{\bibfnamefont{M.}~\bibnamefont{Gilowski}} \bibnamefont{{\it
  et~al.}}, \bibinfo{volume}{quant-ph/0709.1424v1}
  (\bibinfo{year}{2007}).

\bibitem[{\citenamefont{Weiner}(2000)}]{weiner00}
\bibinfo{author}{\bibfnamefont{A.~M.} \bibnamefont{Weiner}},
  \bibinfo{journal}{Rev. Sci. Instr.} \textbf{\bibinfo{volume}{71}},
  \bibinfo{pages}{1929} (\bibinfo{year}{2000}).

\bibitem[{\citenamefont{Meshulach and Silberberg}(1998)}]{silberberg98}
\bibinfo{author}{\bibfnamefont{D.}~\bibnamefont{Meshulach}} \bibnamefont{{\it
  et~al.}}, \bibinfo{journal}{Nature} \textbf{\bibinfo{volume}{396}},
  \bibinfo{pages}{239} (\bibinfo{year}{1998});
\bibinfo{author}{\bibfnamefont{N.}~\bibnamefont{Dudovich}} \bibnamefont{{\it
  et~al.}}, \bibinfo{journal}{Phys. Rev. Lett.} \textbf{\bibinfo{volume}{86}},
  \bibinfo{pages}{47} (\bibinfo{year}{2001});
\bibinfo{author}{\bibfnamefont{J.}~\bibnamefont{Degert}} \bibnamefont{{\it
  et~al.}}, \bibinfo{journal}{Phys. Rev. Lett.} \textbf{\bibinfo{volume}{89}},
  \bibinfo{pages}{203003} (\bibinfo{year}{2002});
\bibinfo{author}{\bibfnamefont{B.}~\bibnamefont{Chatel}} \bibnamefont{{\it
  et~al.}}, \bibinfo{journal}{Phys. Rev. A} \textbf{\bibinfo{volume}{68}},
  \bibinfo{pages}{041402} (\bibinfo{year}{2003});
\bibinfo{author}{\bibfnamefont{A.}~\bibnamefont{Monmayrant}} \bibnamefont{{\it
  et~al.}}, \bibinfo{journal}{Phys. Rev. Lett.} \textbf{\bibinfo{volume}{96}},
  \bibinfo{pages}{103002} (\bibinfo{year}{2006}{\natexlab{a}}); \bibinfo{journal}{Opt. Lett.} \textbf{\bibinfo{volume}{31}},
  \bibinfo{pages}{410} (\bibinfo{year}{2006}{\natexlab{b}});  \bibinfo{journal}{Opt. Commun.} \textbf{\bibinfo{volume}{264}},
  \bibinfo{pages}{256} (\bibinfo{year}{2006}{\natexlab{c}}).
\bibinfo{author}{\bibfnamefont{A.}~\bibnamefont{Assion}} \bibnamefont{{\it
  et~al.}}, \bibinfo{journal}{Science} \textbf{\bibinfo{volume}{282}},
  \bibinfo{pages}{919} (\bibinfo{year}{1998}).
\bibinfo{author}{\bibfnamefont{C.}~\bibnamefont{Daniel}} \bibnamefont{{\it
  et~al.}}, \bibinfo{journal}{Science} \textbf{\bibinfo{volume}{299}},
  \bibinfo{pages}{536} (\bibinfo{year}{2003}).
\bibinfo{author}{\bibfnamefont{K.}~\bibnamefont{Ohmori}} \bibnamefont{{\it
  et~al.}},  \bibinfo{journal}{Phys. Rev. Lett.} \textbf{\bibinfo{volume}{91}},
  \bibinfo{pages}{243003} (\bibinfo{year}{2003}); \textbf{\bibinfo{volume}{96}},
  \bibinfo{pages}{093002} (\bibinfo{year}{2006}).
\bibinfo{author}{\bibfnamefont{H.}~\bibnamefont{Katsuki}} \bibnamefont{{\it et~al.}}, \bibinfo{journal}{Science} \textbf{\bibinfo{volume}{311}}, \bibinfo{pages}{1589} (\bibinfo{year}{2006}).



\bibitem[{\citenamefont{Ahn {\it et~al.}}(2000)\citenamefont{Ahn, Weinacht, and
  Bucksbaum}}]{BucksbaumScience00}
\bibinfo{author}{\bibfnamefont{J.}~\bibnamefont{Ahn}} \bibnamefont{{\it
  et~al.}}, \bibinfo{journal}{Science} \textbf{\bibinfo{volume}{287}},
  \bibinfo{pages}{463} (\bibinfo{year}{2000}).

\bibitem[{\citenamefont{Zheng and Weiner}(2000)}]{Weiner00SHG}
\bibinfo{author}{\bibfnamefont{Z.}~\bibnamefont{Zheng}} \bibnamefont{{\it
  et~al.}}, \bibinfo{journal}{Opt. Lett.} \textbf{\bibinfo{volume}{25}},
  \bibinfo{pages}{984} (\bibinfo{year}{2000}).

\bibitem[{\citenamefont{Amitay {\it et~al.}}(2002)\citenamefont{Amitay,
  Kosloff, and Leone}}]{AMitay02ANDgate}
\bibinfo{author}{\bibfnamefont{Z.}~\bibnamefont{Amitay}} \bibnamefont{{\it
  et~al.}}, \bibinfo{journal}{Chem. Phys. Lett.}
  \textbf{\bibinfo{volume}{359}}, \bibinfo{pages}{8} (\bibinfo{year}{2002}).

\bibitem[{\citenamefont{Bhattacharya {\it
  et~al.}}(2002)\citenamefont{Bhattacharya, van Linden van~den Heuvell, and
  Spreeuw}}]{Spreeuw_PRL02}
\bibinfo{author}{\bibfnamefont{N.}~\bibnamefont{Bhattacharya}}
  \bibnamefont{{\it et~al.}}, \bibinfo{journal}{Phys. Rev. Lett.}
  \textbf{\bibinfo{volume}{88}}, \bibinfo{pages}{137901}
  (\bibinfo{year}{2002}).

\bibitem[{\citenamefont{Brainis {\it et~al.}}(2003)\citenamefont{Brainis,
  Lamoureux, Cerf, Emplit, Haelterman, and Massar}}]{Massar_PRL03}
\bibinfo{author}{\bibfnamefont{E.}~\bibnamefont{Brainis}} \bibnamefont{{\it
  et~al.}}, \bibinfo{journal}{Phys. Rev. Lett.} \textbf{\bibinfo{volume}{90}},
  \bibinfo{pages}{157902} (\bibinfo{year}{2003}).

\bibitem[{\citenamefont{Londero {\it et~al.}}(2004)\citenamefont{Londero,
  Dorrer, Anderson, Wallentowitz, Banaszek, and Walmsley}}]{Walmsley_PRA04}
\bibinfo{author}{\bibfnamefont{P.}~\bibnamefont{Londero}} \bibnamefont{{\it
  et~al.}}, \bibinfo{journal}{Phys. Rev. A} \textbf{\bibinfo{volume}{69}},
  \bibinfo{pages}{010302} (\bibinfo{year}{2004}).

\bibitem[{Mor()}]{Morgner}
\bibinfo{note}{For an instructive explanation of the temporal Talbot effect see
  F. Mitschke and U. Morgner, Optics and Photonics News \textbf{9} (6), 45
  (1998).}

\bibitem[{\citenamefont{Monmayrant and Chatel}(2004)}]{pulseshaperRSI04}
\bibinfo{author}{\bibfnamefont{A.}~\bibnamefont{Monmayrant}} \bibnamefont{{\it
  et~al.}}, \bibinfo{journal}{Rev. Sci. Instr.} \textbf{\bibinfo{volume}{75}},
  \bibinfo{pages}{2668} (\bibinfo{year}{2004}).

\bibitem[{\citenamefont{Wang {\it et~al.}}(2001)\citenamefont{Wang, Zheng,
  Leaird, Weiner, Dorschner, Fijol, Friedman, Nguyen, and
  Palmaccio}}]{wang_shaping2001}
\bibinfo{author}{\bibfnamefont{H.}~\bibnamefont{Wang}} \bibnamefont{{\it
  et~al.}}, \bibinfo{journal}{IEEE J. Sel. Top. Quantum Electron.}
  \textbf{\bibinfo{volume}{7}}, \bibinfo{pages}{718} (\bibinfo{year}{2001}).

\bibitem[{\citenamefont{Vaughan {\it et~al.}}(2006)\citenamefont{Vaughan,
  Feurer, Stone, and Nelson}}]{Vaughan_optexp2006}
\bibinfo{author}{\bibfnamefont{J.}~\bibnamefont{Vaughan}} \bibnamefont{{\it
  et~al.}}, \bibinfo{journal}{Optics Express} \textbf{\bibinfo{volume}{14}},
  \bibinfo{pages}{1314} (\bibinfo{year}{2006}).

\bibitem[{\citenamefont{Wohlleben {\it et~al.}}(2004)\citenamefont{Wohlleben,
  Degert, Monmayrant, Chatel, Motzkus, and Girard}}]{RbShapingAPB04}
\bibinfo{author}{\bibfnamefont{W.}~\bibnamefont{Wohlleben}} \bibnamefont{{\it
  et~al.}}, \bibinfo{journal}{Appl. Phys. B} \textbf{\bibinfo{volume}{79}},
  \bibinfo{pages}{435 } (\bibinfo{year}{2004}).

\bibitem[{\citenamefont{\v{S}tefa\v{n}\'{a}k {\it
  et~al.}}(2007)\citenamefont{\v{S}tefa\v{n}\'{a}k, Merkel, Schleich, Haase,
  and Maier}}]{Schleich07truncation}
\bibinfo{author}{\bibfnamefont{M.}~\bibnamefont{\v{S}tefa\v{n}\'{a}k}}
  \bibnamefont{{\it et~al.}}, \bibinfo{journal}{New J. Phys.}
  \textbf{\bibinfo{volume}{9}}  \bibinfo{pages}{370 } (\bibinfo{year}{2007}).

\bibitem[{\citenamefont{Merkel {\it et~al.}}(2007)\citenamefont{Merkel,
  Schleich, Sh.~Averbukh, and Girard}}]{Merkel07PRA2}
\bibinfo{author}{\bibfnamefont{W.}~\bibnamefont{Merkel}} \bibnamefont{{\it
  et~al.}}, \bibinfo{journal}{Phys. Rev. A} \textbf{\bibinfo{volume}{{\it in
  preparation}}} (\bibinfo{year}{2007}).

\bibitem[{\citenamefont{Agarwal}(1995)}]{Agarwal95}
\bibinfo{author}{\bibfnamefont{G.~S.} \bibnamefont{Agarwal}},
  \bibinfo{journal}{Opt. Commun.} \textbf{\bibinfo{volume}{119}},
  \bibinfo{pages}{30} (\bibinfo{year}{1995}).

\bibitem[{\citenamefont{Friesch {\it et~al.}}(2000)\citenamefont{Friesch,
  Marzoli, and Schleich}}]{Schleich00}
\bibinfo{author}{\bibfnamefont{O.~M.} \bibnamefont{Friesch}} \bibnamefont{{\it
  et~al.}}, \bibinfo{journal}{New J. Phys.} \textbf{\bibinfo{volume}{2}},
  \bibinfo{pages}{1} (\bibinfo{year}{2000}).

\bibitem[{\citenamefont{St\"{u}tzle {\it
  et~al.}}(2005)\citenamefont{St\"{u}tzle, Gobel, Horner, Kierig, Mourachko,
  Oberthaler, Efremov, Fedorov, Yakovlev, van Leeuwen {\it
  et~al.}}}]{Schleich05}
\bibinfo{author}{\bibfnamefont{R.}~\bibnamefont{St\"{u}tzle}} \bibnamefont{{\it
  et~al.}}, \bibinfo{journal}{Phys. Rev. Lett.} \textbf{\bibinfo{volume}{95}},
  \bibinfo{pages}{110405} (\bibinfo{year}{2005}).

\bibitem[{Fac()}]{Factor2}
\bibinfo{note}{For an introduction to atom optics in quantized light fields see
  Chapt. 20 in: W. P. Schleich, {\it Quantum optics in phase space} (Wiley-VCH,
  Berlin, 2001)}.

\end{thebibliography}
\bibliographystyle{apsrev}


\end{document}